# Breakdown of statistical inference from some random experiments

**Authors:** Marian Kupczynski[1] and Hans De Raedt[2]

**Affiliations:**

[1] Département de l'Informatique, Université du Québec en Outaouais (UQO), Case postale 1250, succursale Hull, Gatineau. Quebec, J8X 3X 7, Canada

[2] Zernike Institute for Advanced Materials, University of Groningen, Nijenborgh 4,
NL-9747 AG Groningen, The Netherlands

Correspondence to: h.a.de.raedt@rug.nl

**Abstract:** Many experiments can be interpreted in terms of random processes operating according to some internal protocols. When experiments are costly or cannot be repeated only one or a few finite samples are available. In this paper we study data generated by pseudo-random computer experiments operating according to particular internal protocols. We show that the standard statistical analysis performed on a sample, containing $10^5$ data points or more, may sometimes be highly misleading and statistical errors largely underestimated. Our results confirm in a dramatic way the dangers of standard asymptotic statistical inference if a sample is not homogenous. We demonstrate that analyzing various subdivisions of samples by multiple chi-square tests and chi-square frequency graphs is very effective in detecting sample inhomogeneity. Therefore to assure correctness of the statistical inference the above mentioned chi-square tests and other non-parametric sample homogeneity tests should be incorporated in any statistical analysis of experimental data. If such tests are not performed the reported conclusions and estimates of the errors cannot be trusted.

## 1 Introduction

Outcomes of experiments or surveys in various domains of science are usually interpreted as observed values of one or more random variables obeying some, in general, multivariate probability distribution. Gathered data are often assumed to be simple random samples. A random sample is simple if it is homogeneous and all trials are independent.

The dangers of statistical inference based on finite samples are well known to statisticians but many experimentalists seem to be unaware of them. Let us cite here (Kruskal 1988) : "incorrect assumptions of 'simple' random sampling can invalidate statistical inference" .

Computer packages for statistical analysis produce descriptive statistics and outcomes of various significance tests. However these packages cannot replace statistical thinking and mistaken conclusions are often drawn in a variety of studies because the researchers do not appreciate the significance of the assumptions about the probability distribution underlying a model and for other reasons (Hand 2015).

Particular caution is needed in the case where only one large sample of data is available and we want to make a sound statistical inference based on it, as in for example, the data obtained in the experiments of (Christensen et al 2013) and (Giustina et al 2013). One may not simply assume that the experimental data are 'simple' random samples without verifying it.

Many experimentalists believe, when a sample size is $10^4$ or larger, that a sample average and a sample mean error give reliable information about studied statistical population even if a studied sample is not a perfect *simple random sample.* In this paper we show that such belief is unjustified and a careful study of sample homogeneity is always necessary. Some experimental devices, operating according to specific internal protocols may produce strange, but legitimate, outcomes which usually would be considered as outliers and rejected.

In order to explore possible anomalies in large finite samples, we study several pseudo-random computer experiments generating time series of discrete data according to different internal protocols. We use the term "internal" to indicate that the details of the protocol are inaccessible to any person analyzing the data, as in real-life applications. We demonstrate that standard statistical inference of one or a few of large samples (containing as much as $10^7$ data points) generated by some of these protocols in terms of standard errors and various confidence intervals can be highly misleading.

By subdividing our samples into 100 bins (each bin containing $10^5$ data items) and by performing 4950 chi-square bin-to-bin compatibility tests we demonstrate that samples produced by some of our computer experiments are not homogenous, explaining the invalid conclusion based on the performed significance tests. For the samples which are homogenous we obtain close to perfect agreement with a corresponding probabilistic model.

In our paper we not only demonstrate the dramatic consequences of sample inhomogeneity but we suggest which preliminary supplementary statistical tests of the data should be performed in order to assure a sound statistical inference. These tests detected the anomalies in our computer generated samples without making use of any knowledge about a particular protocol.

## 2 Standard statistical inferences

Let us assume that $A$ can take $k$ different values: $a_1, a_2, \ldots a_k$. In a long run of the experiment we obtain a random sample $S=\{x_1, x_2, \ldots x_N\}$ of size N which according to standard sampling methods is interpreted as an observation of a multivariate random variable $\{A_1, A_2, \ldots A_N\}$ where $A_i$ are independent and identically distributed random variables (i.i.d.): $A_i \sim D$.

The empirical frequency distributions of various outcomes $f_i = \#(x_j=a_i)/N$ are found and believed to approach the probabilities provided by the theory. Furthermore the probability distribution of the variable $\bar{A} = \frac{1}{N}\sum_{i=1}^{N} A_i$ is, due to the central limit theorem (CLT), believed to be well approximated by a normal distribution $N(\mu_{\bar{A}}, \sigma_{\bar{A}}^2)$ with $\mu_{\bar{A}} = \mu_A$ and $\sigma_{\bar{A}} = \sigma_A/\sqrt{N}$ where $\mu_A$ and $\sigma_A$ are the mean and the standard deviation of the random variable *A*. In spite of the fact that CLT is valid when N tends to infinity it is often assumed that already for N ≥ 30 the normal distribution provides a reasonable approximation and that the unknown variance $\sigma_A^2$ can be replaced by a value of its unbiased estimator $s^2$:

$$s^2 = \frac{\sum_{i=1}^{N}(x_i - \bar{x})^2}{N-1} \qquad (1)$$

A sample mean $\bar{x}$ is considered as a good estimate of $<A> = \mu_A$ and a standard error of the mean SEM= $s/\sqrt{N}$ as a good estimate of $\sigma_{\bar{A}} = \sigma_A/\sqrt{N}$.

As *N* increases the confidence in the validity of the approximation by a normal distribution is increasing and the errors become smaller and smaller. The most exact probabilistic statement, if the normal-distribution approximation is valid, can be expressed in terms of the confidence intervals $I_\alpha$ :

$$I_\alpha = \left[\bar{x} - z_{\alpha/2} s/\sqrt{N},\ \bar{x} + z_{\alpha/2} s/\sqrt{N}\right] \qquad (2)$$

saying that the probability that the interval $I_\alpha$ covers the unknown value $\mu_A$ is $(1-\alpha)$.

If the asymptotic normality of the distribution is not assumed, the Chebyshev's inequality can used and the confidence interval (2) is replaced by:

$$I_c = \left[\bar{x} - cs/\sqrt{N},\ \bar{x} + cs/\sqrt{N}\right] \qquad (3)$$

and the probability that the interval $I_c$ covers the unknown value $\mu_A$ is $(1-\frac{1}{c^2})$.

Of course the estimation of SEM= $s/\sqrt{N}$ is valid if the variables $A_i$ are independent and identically distributed random variables (i.i.d.): $A_i \sim D$. But this has to be carefully checked and not taken for granted. The Chebychev's inequality for the finite samples is valid under the supplementary assumptions (Saw et al. 1984) and (Kaban 2012).

**3 Experiments and invisible internal protocols**

Let us imagine a following experiment. A signal is entering a measuring device (considered to be a black box) and from time to time some discrete outcomes are produced and a sample S is obtained. If the outcomes seem to be randomly distributed we could assume a following probabilistic model:

- a signal is described by a probability distribution $p_1(m)$
- a state of the device at the moment of a measurement is described by a probability distribution $p_2(n)$
- the output of the device is one of discrete values $A(m, n)$.

If this simple probabilistic model is assumed then the expectation value:

$$\langle A \rangle = \sum_{m,n} A(m,n) p_1(m) p_2(n) \qquad (4)$$

The probability distribution $p(A(m, n)=a)$ and the standard deviation $\sigma_A$ are easily found and compared with experimental data.

As we mentioned above we do not know how our device produces successive outcomes. Therefore we perform several Monte Carlo simulations using various possible internal protocols and we compare the properties of the finite samples generated by these protocols.

We define three different protocols:

Protocol 1= $(N_1, 1, m)$ :

- generate one value of *n* and one value of *m* using $p_1(m)$ and $p_2(n)$
- evaluate A(*m*, *n*) and output this value
- repeat the process $N_1$ times in order to create a sample of size $N=N_1$.

Protocol 2= $(N_1, N_2, m)$ :

- generate one value of *n* and $N_2 > 1$ values of *m* using $p_1(m)$ and $p_2(n)$
- evaluate *A(m, n)* and output the values for the $N_2$ different values of *m*
- repeat the process $N_1$ times in order to create a sample of a size $N=N_1 N_2$.

Protocol 3= $(N_1, N_2, n)$ :

- generate one value of *m* and $N_2 > 1$ values of *n* using $p_1(m)$ and $p_2(n)$
- evaluate *A(m, n)* and output the values for the $N_2$ different values of *n*
- repeat the process $N_1$ times in order to create a sample of a size $N=N_1 N_2$

From the created finite samples we compute the frequency distributions, sample means and a sample standard deviations. In the limit when both $N_1$ and $N_2$ tend to infinity one might expect that estimates of proportions and averages should be consistent for all different protocols.

However, we find significant differences between our large samples. These are due to the differences between $p_1(m)$ and $p_2(n)$, asymmetric sampling of *n* and *m* and the asymmetry of the function A (*m, n*) in one of our two models.

**4 Models and Monte Carlo simulation**

We perform several Monte Carlo simulations based on two different models.

**Model 1**.

We choose *A(m, n)*=((*m*+*n*+1)mod 3)+1 where *m* and *n* are random variables taking values 0, 1, or 2 and 0 or 1, respectively. The probability distributions are defined as:

$$p_1(m): \quad p_1(0) = 1/8, \; p_1(1) = 1/2 \text{ and } p_1(2) = 3/8 \tag{5}$$

$$p_2(n): \quad p_2(0) = 1/4 \text{ and } p_2(1) = 3/4 \tag{6}$$

This device produces three outcomes 1, 2 or 3 distributed as : p(A=1)=15/32, p(A=2)=10/32 and p(A=3)=7/32. From (1) we find that the expectation value $\langle A \rangle$ =1.75.

**Model 2** :

We choose $A(m, n) = (m+2n)^2$ where $m$ = 40, 80, or 100 and $n$ = 100 or 500. The probabilities $p_1(m)$ and $p_2(n)$ have the same values and are assigned in the same order as in (5) and (6).

This device produces 6 outcomes denoted by a corresponding couple (m, n): (40,100)=57600, (40,500)=1081600, (80,100)=78400, (80,500)=1166400, (100,100)=90000 and (100,500)=1210000.

Corresponding probability distribution is defined as: p(40,100)=1/32, p(40,500)=3/32, p(80,100)=1/8, p(80,500)=3/8, p(100,100)=9/32, p(490000)=9/32 and the theoretical expectation value calculated using (4) is $\langle A \rangle$ = 899150.

It is convenient to use normalized averages $\langle A \rangle_S / \langle A \rangle$, where $\langle A \rangle_S$ is a sample average and $\langle A \rangle$ is a theoretical expectation value found using the probabilistic model (4). Using the three different protocols described above, 100 large samples of the same size were generated for each protocol. The output of our computer program contains among others (see Appendix for a representative output):

- for the runs labeled 1, 25, 50, 75 and 100: standard errors of the mean (SEM) and SEM calculated using 5, 10, 100 bins.
- $\langle A \rangle_S / \langle A \rangle$ and corresponding SEM obtained by using all the data from 100 runs.
- the averages and the maximum values of $\chi^2$ and the smallest P-values obtained from 99x50 chi-square cross-comparison of all 100 runs produced by each of the protocols
- for each protocol, a histogram of the $\chi^2$ values obtained from 99x50 chi-square tests.

## 5 Experimental results and data analysis

We created samples containing $10^4$ and $10^5$ data items by choosing $N_1$=4 or 40 and $N_2$=250, 2500 and 25000 or vice-versa. By repeating the computer experiments 100 times, we generated large random samples containing $10^6$ or $10^7$ outcomes subdivided into 100 bins. We have checked that our conclusions did not depend on the particular random number generator used and that they did not change when we repeated the experiments.

First we want to test the hypothesis $H_0$: $\langle A \rangle_S / \langle A \rangle \geq 1$ using the data generated in a run 25 by our 3 protocols for $N_1$=4, $N_2$=2500 and the model 1.

**Table 1.** Statistical inference based on a single runs and on the collection of 100 runs.

| Protocol | $\langle A \rangle_S / \langle A \rangle$ | SEM | 99.9% CI | 99% CI (Cheb.) | $\langle A \rangle_S / \langle A \rangle$ 100 runs |
|---|---|---|---|---|---|
| 1 | 0.9887 | $0.4491 \times 10^{-2}$ | [0.9740, 1.0034] | [0.9438, 1.0336] | $0.9997 \pm 0.4346 \times 10^{-3}$ |
| 2 | 0.8527 | $0.4923 \times 10^{-2}$ | [0.8365, 0.8689] | [0.8035, 0.9019] | $0.9833 \pm 0.1127 \times 10^{-1}$ |
| 3 | 0.9236 | $0.3941 \times 10^{-2}$ | [0.9106, 0.9366] | [0.8842, 0.9630] | $0.9994 \pm 0.6493 \times 10^{-2}$ |

In Table 1, 99% CI (Cheb.) is the confidence interval based on the Chebyshev's inequality. As in (Khrennikov et al 2014) we use (3) with c=10. The 99% CI (Cheb.) corresponds to 10 standard deviations confidence interval. To find 99.9% CI we use (2) with $z_{\alpha/2} = 3.29$.

Since the averages shown in Table 1 were calculated using $10^4$ data points and the confidence intervals for the different protocols do not overlap one might conclude that three samples were drawn from different statistical populations.

The data of protocol 1 is in concert with hypothesis $H_0$ but one can with great confidence reject $H_0$ based on the data of protocols 2 and 3. Obviously, this conclusion would be incorrect as the averages obtained from 100 independent runs are consistent with the expected theoretical value of $\langle A \rangle_S / \langle A \rangle$ =1.

Analyzing in the same manner the results for runs 1, 50, 75 and 100, as we did for the run 25, we find that all the 5 CIs cover the correct value $\langle A \rangle_S / \langle A \rangle$ =1 in the case of the protocol 1, in contrast to 2 out of 5 CIs for the protocols 2 and 3. Using the 99.9% CI, we expect that only 1 out of 1000 cases may not include the correct value. Using the 99% CI (Cheb.), we expect that 1 out of 100 intervals may not include the correct value.

Therefore, based on the data of one long experimental run only, the use of the CLT or Chebyshev's inequality and related confidence intervals does not guarantee the correctness of the statistical inference and a more detailed analysis is required.

Let us present now another example showing an even more dramatic breakdown of standard statistical inference if used to test hypothesis $H_0$: $\langle A \rangle_S / \langle A \rangle \leq 1$. We consider the data of 4 runs generated by model 2 and protocol 3 with $N_1$= 4, $N_2$ =25000, each containing $10^5$ data.

**Table 2.** Testing $H_0$: $\langle A \rangle_S / \langle A \rangle \leq 1$ using model 2 and 4 runs produced by the protocol 3.

| Run | 1 | 2 | 3 | 4 | 100 runs |
|---|---|---|---|---|---|
| $\langle A \rangle_S / \langle A \rangle$ | 0.3928 | $0.1304 \times 10^{+1}$ | $0.1304 \times 10^{+1}$ | $0.1303 \times 10^{+1}$ | 0.9727 |
| SEM | $0.1665 \times 10^{-2}$ | $0.1396 \times 10^{-3}$ | $0.1395 \times 10^{-3}$ | $0.1397 \times 10^{-3}$ | $0.2851 \times 10^{-1}$ |
| $1 - \langle A \rangle_S / \langle A \rangle$ | +364 SEM | -2177 SEM | -2236 SEM | -2168 SEM | +0.95 SEM |

Once again the use of confidence intervals may lead to incorrect conclusions. From Chebyshev's inequality, the probability of observing a 2000 SEM deviation from zero is $0.25 \times 10^{-6}$. Therefore, if only runs 1-4 were available one would with great confidence conclude using the runs 2-4 that $1 - \langle A \rangle_S / \langle A \rangle$ was negative and reject the null hypothesis $H_0$. The run 1 would be considered as an outlier.

The calculation of SEM= $s/\sqrt{N}$ assumes the independence of sampling distributions. If these distributions are not independent the SEM might be larger. For example if X and Y are independent random variables and var(X) = var(Y) then var (X+Y)= 2 var(X) but if X=Y then var (2X) = 4

var(X). Even if we replaced the SEM by *s*, the data in the columns 2-4 would show at least a 6 standard deviation violation of the tested inequality.

Being more cautious, one could divide the runs into 5, 10 and 100 bins and estimate the SEM using the binned data and then check the consistency of successive bins by using a series of chi-square tests. For run 1 we obtain SEM= 0.1057 using 10 bins and SEM=0.04836 when using 100 bins. Moreover, the chi-square tests for 5 and 10 bins of the run 1 show large bin-to bin variability yielding a minimal P-value = 0 (meaning that the numerical value is smaller $10^{-300}$), providing additional justification for the rejection of the run 1. Repeating the same analysis for the runs 2, 3 and 4 the bin-to-bin consistency of the data produced by these runs is confirmed.

Since the value of $\langle A \rangle_S / \langle A \rangle$ obtained by averaging the data of 100 runs (column 5 of Table 2) does not allow to reject $H_0$ the preceding statistical inference based on only four long runs was highly misleading.

## 6 Empirical frequencies

One might expect that in spite of the run-to-run variability, the empirical frequency distributions of outcomes averaged over all 100 runs (i.e. obtained from samples of size $10^7$) should be consistent with the theoretical predictions. However, this expectation is not supported by the data. As the standard error $\hat{\sigma}_{\hat{p}_1 - \hat{p}_2}$ of the difference of two proportion estimators is smaller than $(2n)^{-0.5}$ ( $(2n)^{-0.5}$=0.002 for n=$10^7$), we might detect differences of more than $9\hat{\sigma}_{\hat{p}_1 - \hat{p}_2}$ between the observed frequencies.

In particular for model 2 and ($N_1$= 4, $N_2$=25000), the observed frequencies for the protocol 1 and 3 together with the theoretical predictions from (4-6) for model 2 are displayed in Table 3.

From Table 3 it is clear that the data produced using the protocol 3 deviate significantly from the multinomial distribution. Note the close-to-perfect agreement between the data of protocol 1 and the theoretical model.

**Table 3.** Comparison of observed frequencies for model 2 and ($N_1$= 4, $N_2$=25000)

| (m, n)         | (40,100) | (40,500) | (80,100) | (80,500) | (100,100) | (100,500) |
|----------------|----------|----------|----------|----------|-----------|-----------|
| $p_1$(m) $p_1$(n) | 0.03125  | 0.09375  | 0.125    | 0.375    | 0.09375   | 0.28125   |
| Protocol 1     | 0.03127  | 0.09371  | 0.1252   | 0.3749   | 0.09363   | 0.2813    |
| Protocol 3     | 0.0265   | 0.0984   | 0.1063   | 0.3936   | 0.07964   | 0.2955    |

## 7 Chi-square tests and histograms

The statistical inference based on individual runs of both models operating according to the protocol 1 is consistent with the probabilistic model (4-6). In contrast, the statistical inference based on data created according to protocols 2 and 3 turned out to be unreliable, indicating that the large data sets produced by these protocols are not 'simple' random samples. Therefore, to make reliable inferences, it is necessary to test the homogeneity of the samples in more systematic way. A simple, effective procedure is to make chi-square compatibility tests between pairs of bins from various partitions of the samples.

In particular, we concentrate on the analysis of the data obtained in 100 repetitions of our simulation experiments treating individual runs as the bins of large samples containing $10^6$ or $10^7$ data points. To compare these bins we make 99x50=4950 chi-square tests using the statistics:

$$\chi^2 = \sum_{i=1}^{k} \frac{(R_i - S_i)^2}{R_i + S_i} \tag{7}$$

where $R_i$ and $S_i$ are counts of the same outcomes in the compared bins and k=3 and k=6 for model 1 and model 2 respectively. The $\chi^2$ statistics has respectively $\nu = 2$ and $\nu = 5$ degrees of freedom for model 1 and 2. Since the differences of cell frequencies are large we don't need to use the continuity correction. In essence, this chi-square test tells us whether the counts of various outcomes in the different bins of our large samples are similar.

If the samples are produced according to a multinomial distribution, all trials are identical and independent and $\chi^2$ obeys approximately the chi-square distribution. However, if the experiment producing the outcomes is not multinomial, $\chi^2$ not necessarily obeys the chi-square probability distribution and the calculated P-value = $P(\chi^2 \geq \chi^2(\text{observed}))$ should be used with great care.

Nevertheless, the chi-square tests (7) may detect significant differences between the bins. The P-value is calculated as P-value = $Q(\chi^2(\text{observed})/2, \nu/2)$, where $Q(a,x)$ is the incomplete gamma function. In Table 4 we display the minimum P-values obtained in 4950 chi-square tests performed on the samples created by model 2 using the three different protocols and for various choices of $N_1$ and $N_2$.

**Table 4.** Minimum P-values from 4950 chi-square tests probing the sample homogeneity. An entry "0" indicates that the minimum P-value is smaller than $10^{-300}$.

|  | (25000,4) | (2500,40) | (4,25000) | (4, 2500) |
|---|---|---|---|---|
| Protocol 1 | $0.8537 \times 10^{-3}$ | $0.3038 \times 10^{-3}$ | $0.1970 \times 10^{-3}$ | $0.1029 \times 10^{-2}$ |
| Protocol 2 | $0.2319 \times 10^{-13}$ | $0.8788 \times 10^{-110}$ | 0 | 0 |
| Protocol 3 | $0.1417 \times 10^{-12}$ | $0.5400 \times 10^{-160}$ | 0 | 0 |

The minimum P-value in Table 4 is the value corresponding to the largest $\chi^2$ (observed) in 4950 comparisons. For the protocols 2 and 3 the probability P of observing at least one so large value $\chi^2$ (observed), in 4950 chi–square comparison tests, can be very conservatively estimated using the Bonferroni correction : P ~ 4950 x (min P-value). We see that if we multiply the rows 3 and 4 by 4950 the entries remain still negligible.

It then follows that we have no reasons to doubt the homogeneity of the samples generated by using the protocol 1. The samples created by the protocols 2 and 3 are not homogenous. It can be clearly seen from histograms of all 4950 observed values of $\chi^2$ displayed in Figs.1 and 2 below.

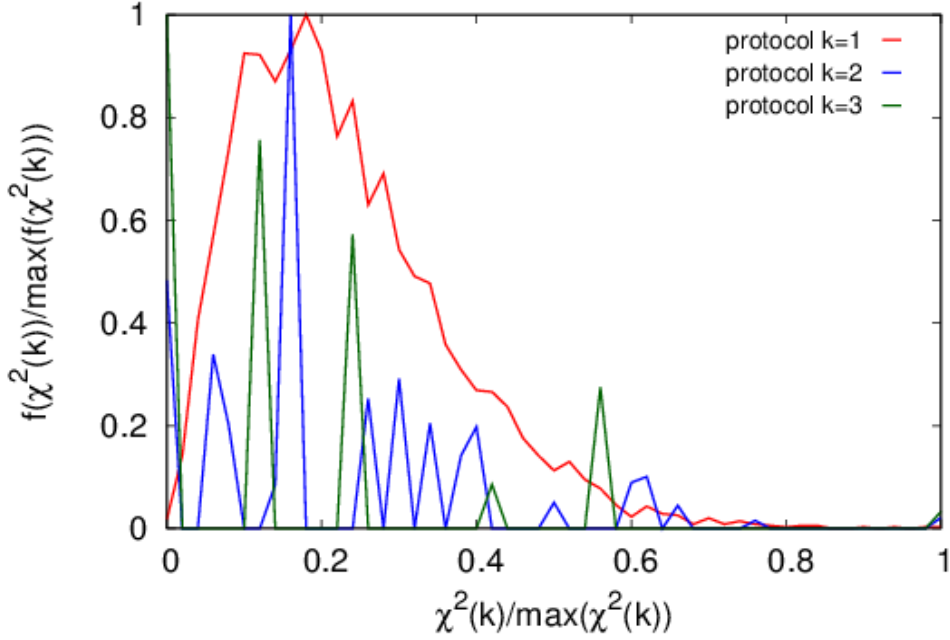

FIG. 1: Relative frequency $f(\chi^2(k))/\max f(\chi^2(k)))$ of observing $\chi^2(k)$, versus $\chi^2(k)/\max \chi^2(k)$ as obtained for protocols $k=1,2,3$, ($N_1=4, N_2=2500$) and model 2 and 100 repetitions of the experiment. Note that the values of max $\chi^2(k)$ may vary significantly with the protocol k: we have max $\chi^2(1) = 0.2045 \times 10^2$, max $\chi^2(2) = 0.2 \times 10^5$, and max $\chi^2(3) = 0.12 \times 10^5$.

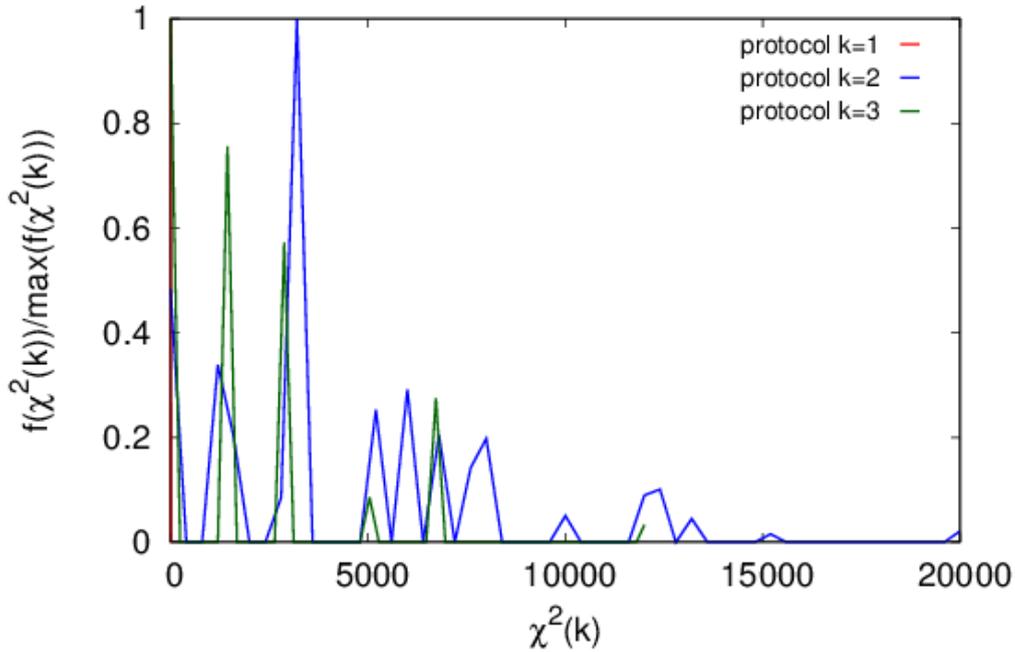

FIG. 2: Relative frequency of $\chi^2(k)$ versus $\chi^2(k)$ for protocols $k=1,2,3$ as obtained for protocols $k=1,2,3$, ($N_1=4, N_2=2500$) and model 2 and 100 repetitions of the experiment.

# 8 Conclusions

We simulated several pseudo-random computer experiments producing 3 or 6 different discrete outcomes. We generated large samples of sizes $10^4$ - $10^7$ and observed a dramatic breakdown of standard statistical inference. This breakdown was due to the fact that some samples were not homogeneous. We demonstrated this by using various partitions of the samples, bin-to-bin chi-square tests and observed $\chi^2$ frequency histograms.

In general, this bin-to-bin chi-square tests test is easy to implement, see Press et al. (2003), easy to use and, as shown in the present paper, allows detecting anomalies in experimental or computer generated samples very effectively. Therefore, we suggest that this should be the first test for homogeneity of the data sample. The procedure is straightforward. Suppose that the data set S consists of *N* items and that a unique label 1,…,*k* has been assigned to each item. In practice such an assignment can always be made. The number *k* defines the number of different bins and should be much smaller than the total number items *N*. The next step is to partition the whole data set into a number *M* of smaller sets $s_1,…,s_M$. The number *M* should be chosen such that *N/M* is large compared to the number of different bins *k* such that for each set $s_1,…,s_M$ the number of items per bin is a reasonably large number, not just zero or one. The final step is then to compute $\chi$ according to Eq.(7) with $(R_i,S_i)$ being all possible pairs $(s_j,s_{j'})$ with $1 \leq j < j' \leq M$. The procedure just sketched uses fragments of the data sets to perform the test but it some cases, one may want to test if the distribution of items complies with a given distribution. In such a case, one can use the same procedure to compare the $(R_i,S_i)$ where the former are taken from the data set and the latter is taken from data generated according to the given distribution, see Press et al. (2003) for more details.

Once the anomalies are detected one has at disposal several non-parametric comparison tests [see for example, Lehmann (2006) and Corder and Foreman (2014) and several specific tests invented to study time series of data [see for example, Box et al. (2008] which can be used to obtain more detailed information about experimental data.

Summarizing: as the deviations from homogeneity can invalidate the statistical inference homogeneity tests should become a standard part of statistical analysis of any large sample of experimental data in any domain of science.

# References

Kruskal, W.: Miracles and Statistics: The Casual Assumption of Independence, Journal of the American Statistical Association 83, 929–940 (1988)

Hand, D.J.: Statistics and computing : the genesis of data science, *Stat Comput* **25**, 705–711 (2015)

Christensen B.G. et al.: Detection -Loophole-Free Test of Quantum Nonlocality, and Applications, PRL **111**, 1304-06 (2013)

Giustina M. et al.: Bell violation with entangled photons, free of the fair-sampling assumption, Nature **497**, 227-230 (2013).

Khrennikov A. et al.:On the equivalence of Clauser-Horne and Eberhard inequality based tests, Phys. Scr. **T163** (2014) 014019 (8pp) *arXiv:1403.2811v2* [quant-phys] (2014)

Saw, J.G. and M.C.K. Yang M.C.K. and   Mo T.C.: Chebyshev Inequality with Estimated Mean and Variance, The American Statistician 38 (2), 130–132 (1984)

Kabán, A.: Non-parametric detection of meaningless distances in high dimensional data,  Statistics and Computing **22** , 375–385 (2012)

Lehmann E.L.:Nonparametrics: Statistical Methods Based on Ranks, Springer, New York (2006)

Corder G.W. and   Foreman D.I.: Nonparametric Statistics: A Step-by-Step Approach , Wiley, New York  (2014)

Box G.E.P, Jenkins G.M.and Reinsel G.C.: Time Series Analysis Forecasting and Control, Wiley, Hoboken (2008)

Press, W. H., Flannery, B. P., Teukolsky, S. A.and Vetterling, W. T.: Numerical Recipes, Cambridge University Press (2003)

Appendix: Representative results of the computer simulation

```
   N1 (ntrials)  = 4
   N2 (nsamples) = 2500
   S  (nrepeat)  = 100

   --- A = (Bm(m)+2Bn(n))**2 ---
  Bm :
      0  0.40000E+02
      1  0.80000E+02
      2  0.10000E+03
  Bn :
      0  0.10000E+03
      1  0.50000E+03

  Pm :
      0  0.12500E+00
      1  0.50000E+00
      2  0.37500E+00
  Pn :
      0  0.25000E+00
      1  0.75000E+00
  A, A*Pm*Pn, Pm*Pn  :
      0   0  0.57600E+05  0.18000E+04  0.31250E-01
      0   1  0.10816E+07  0.10140E+06  0.93750E-01
      1   0  0.78400E+05  0.98000E+04  0.12500E+00
      1   1  0.11664E+07  0.43740E+06  0.37500E+00
      2   0  0.90000E+05  0.84375E+04  0.93750E-01
      2   1  0.12100E+07  0.34031E+06  0.28125E+00
  Am :
      0  0.82560E+06
      1  0.89440E+06
      2  0.93000E+06
  An :
      0  0.80150E+05
      1  0.11722E+07
 Outcomes lookup tables
  Number of different outcomes : 6
  P's :  0.31250E-01  0.93750E-01  0.12500E+00  0.37500E+00  0.93750E-01  0.28125E+00
  A's :  0.57600E+05  0.10816E+07  0.78400E+05  0.11664E+07  0.90000E+05  0.12100E+07
  i's :            0            0            1            1            2            2
  j's :            0            1            0            1            0            1
  outcomes(i,j) :
      0   0   1
      0   1   2
      1   0   3
      1   1   4
      2   0   5
      2   1   6
  <A> =   899150.0000000000
  <A> = <Pn.An> = 899150.0000000000
  <A> = <Pm.Am> = 899150.0000000000

 N1 (samples taken) = 4, N2 (samples to estimate An or Am) = 2500

 === Run number 1, normalized to <A> = 1:
 Protocol 1: <A> =  0.9961E+00, StandardDeviationMean =  0.5298E-02
 Protocol 2: <A> =  0.1007E+01, StandardDeviationMean =  0.5167E-02
 Protocol 3: <A> =  0.9996E+00, StandardDeviationMean =  0.5272E-02
  +++ Analysis of this run using 10 bins
  Protocol 1: <A> =  0.9961E+00, StandardDeviationMean =  0.5606E-02
  Protocol 2: <A> =  0.1007E+01, StandardDeviationMean =  0.6328E-02
  Protocol 3: <A> =  0.9996E+00, StandardDeviationMean =  0.1631E+00
  --- Analysis of this run using 100 bins
  Protocol 1: <A> =  0.9961E+00, StandardDeviationMean =  0.5208E-02
  Protocol 2: <A> =  0.1007E+01, StandardDeviationMean =  0.5332E-02
  Protocol 3: <A> =  0.9996E+00, StandardDeviationMean =  0.5284E-01
  --- khi^2 Analysis of this run using 10 bins ---
               <khi2>      max(khi2)    P-value
  Protocol 1:  0.6504E+01  0.1463E+02  0.1205E-01
  Protocol 2:  0.1008E+01  0.5076E+01  0.4067E+00
  Protocol 3:  0.8175E+03  0.2000E+04  0.0000E+00

 === Run number 25, normalized to <A> = 1:
 Protocol 1: <A> =  0.9948E+00, StandardDeviationMean =  0.5299E-02
```

```
Protocol 2: <A> =  0.9931E+00, StandardDeviationMean =   0.5249E-02
Protocol 3: <A> =  0.1000E+01, StandardDeviationMean =   0.5273E-02
 +++ Analysis of this run using 10 bins
 Protocol 1: <A> =  0.9948E+00, StandardDeviationMean =   0.6133E-02
 Protocol 2: <A> =  0.9931E+00, StandardDeviationMean =   0.5268E-02
 Protocol 3: <A> =  0.1000E+01, StandardDeviationMean =   0.1632E+00
 --- Analysis of this run using 100 bins
 Protocol 1: <A> =  0.9948E+00, StandardDeviationMean =   0.5725E-02
 Protocol 2: <A> =  0.9931E+00, StandardDeviationMean =   0.4799E-02
 Protocol 3: <A> =  0.1000E+01, StandardDeviationMean =   0.5285E-01
 --- khi^2 Analysis of this run using 10 bins ---
              <khi2>    max(khi2)   P-value
 Protocol 1:  0.5136E+01  0.1127E+02  0.4620E-01
 Protocol 2:  0.1019E+01  0.5149E+01  0.3980E+00
 Protocol 3:  0.8154E+03  0.2000E+04  0.0000E+00

=== Run number 50, normalized to <A> = 1:
Protocol 1: <A> =  0.1000E+01, StandardDeviationMean =   0.5267E-02
Protocol 2: <A> =  0.1011E+01, StandardDeviationMean =   0.5341E-02
Protocol 3: <A> =  0.1304E+01, StandardDeviationMean =   0.4372E-03
 +++ Analysis of this run using 10 bins
 Protocol 1: <A> =  0.1000E+01, StandardDeviationMean =   0.4751E-02
 Protocol 2: <A> =  0.1011E+01, StandardDeviationMean =   0.6239E-02
 Protocol 3: <A> =  0.1304E+01, StandardDeviationMean =   0.6836E-03
 --- Analysis of this run using 100 bins
 Protocol 1: <A> =  0.1000E+01, StandardDeviationMean =   0.5090E-02
 Protocol 2: <A> =  0.1011E+01, StandardDeviationMean =   0.5720E-02
 Protocol 3: <A> =  0.1304E+01, StandardDeviationMean =   0.4738E-03
 --- khi^2 Analysis of this run using 10 bins ---
              <khi2>    max(khi2)   P-value
 Protocol 1:  0.5359E+01  0.1522E+02  0.9461E-02
 Protocol 2:  0.9264E+03  0.2000E+04  0.0000E+00
 Protocol 3:  0.3625E+01  0.1652E+02  0.5497E-02

=== Run number 75, normalized to <A> = 1:
Protocol 1: <A> =  0.1010E+01, StandardDeviationMean =   0.5207E-02
Protocol 2: <A> =  0.9872E+00, StandardDeviationMean =   0.5212E-02
Protocol 3: <A> =  0.3929E+00, StandardDeviationMean =   0.5266E-02
 +++ Analysis of this run using 10 bins
 Protocol 1: <A> =  0.1010E+01, StandardDeviationMean =   0.4824E-02
 Protocol 2: <A> =  0.9872E+00, StandardDeviationMean =   0.1457E-01
 Protocol 3: <A> =  0.3929E+00, StandardDeviationMean =   0.1633E+00
 --- Analysis of this run using 100 bins
 Protocol 1: <A> =  0.1010E+01, StandardDeviationMean =   0.4623E-02
 Protocol 2: <A> =  0.9872E+00, StandardDeviationMean =   0.6438E-02
 Protocol 3: <A> =  0.3929E+00, StandardDeviationMean =   0.5287E-01
 --- khi^2 Analysis of this run using 10 bins ---
              <khi2>    max(khi2)   P-value
 Protocol 1:  0.3996E+01  0.1002E+02  0.7481E-01
 Protocol 2:  0.1324E+04  0.2000E+04  0.0000E+00
 Protocol 3:  0.8166E+03  0.2000E+04  0.0000E+00

=== Run number 100, normalized to <A> = 1:
Protocol 1: <A> =  0.9970E+00, StandardDeviationMean =   0.5293E-02
Protocol 2: <A> =  0.9908E+00, StandardDeviationMean =   0.5288E-02
Protocol 3: <A> =  0.6965E+00, StandardDeviationMean =   0.6084E-02
 +++ Analysis of this run using 10 bins
 Protocol 1: <A> =  0.9970E+00, StandardDeviationMean =   0.6019E-02
 Protocol 2: <A> =  0.9908E+00, StandardDeviationMean =   0.1320E-01
 Protocol 3: <A> =  0.6965E+00, StandardDeviationMean =   0.2025E+00
 --- Analysis of this run using 100 bins
 Protocol 1: <A> =  0.9970E+00, StandardDeviationMean =   0.4908E-02
 Protocol 2: <A> =  0.9908E+00, StandardDeviationMean =   0.6815E-02
 Protocol 3: <A> =  0.6965E+00, StandardDeviationMean =   0.6106E-01
 --- khi^2 Analysis of this run using 10 bins ---
              <khi2>    max(khi2)   P-value
 Protocol 1:  0.4016E+01  0.8860E+01  0.1148E+00
 Protocol 2:  0.1324E+04  0.2000E+04  0.0000E+00
 Protocol 3:  0.1112E+04  0.2000E+04  0.0000E+00

--- Comparing protocols bin-wise ---
                             max(khi2)  P-value
 Protocols 1(  1) and 1(100): 0.3518E+01 0.6207E+00
 Protocols 2(  1) and 2(100): 0.1200E+05 0.0000E+00
 Protocols 3(  1) and 3(100): 0.1335E+04 0.1571-285
 Protocols 1(  1) and 2(  1): 0.6681E+04 0.0000E+00
 Protocols 1(  1) and 3(  1): 0.4679E+01 0.4563E+00
```

```
 Protocols 2(  1) and 3(  1):   0.6534E+04 0.0000E+00

 Protocols 1( 25) and 1(  1):   0.3876E+01 0.5674E+00
 Protocols 2( 25) and 2(  1):   0.3569E+01 0.6129E+00
 Protocols 3( 25) and 3(  1):   0.5418E+01 0.3670E+00
 Protocols 1( 25) and 2( 25):   0.6835E+04 0.0000E+00
 Protocols 1( 25) and 3( 25):   0.2802E+01 0.7305E+00
 Protocols 2( 25) and 3( 25):   0.6729E+04 0.0000E+00

 Protocols 1( 50) and 1( 25):   0.3872E+01 0.5679E+00
 Protocols 2( 50) and 2( 25):   0.6667E+04 0.0000E+00
 Protocols 3( 50) and 3( 25):   0.2858E+04 0.0000E+00
 Protocols 1( 50) and 2( 50):   0.1472E+04 0.0000E+00
 Protocols 1( 50) and 3( 50):   0.2850E+04 0.0000E+00
 Protocols 2( 50) and 3( 50):   0.3941E+04 0.0000E+00

 Protocols 1( 75) and 1( 50):   0.2827E+01 0.7267E+00
 Protocols 2( 75) and 2( 50):   0.3334E+04 0.0000E+00
 Protocols 3( 75) and 3( 50):   0.1200E+05 0.0000E+00
 Protocols 1( 75) and 2( 75):   0.6164E+03 0.5785-130
 Protocols 1( 75) and 3( 75):   0.5186E+04 0.0000E+00
 Protocols 2( 75) and 3( 75):   0.5433E+04 0.0000E+00

 Protocols 1(100) and 1( 75):   0.1054E+02 0.6134E-01
 Protocols 2(100) and 2( 75):   0.1668E+04 0.0000E+00
 Protocols 3(100) and 3( 75):   0.1334E+04 0.2499-285
 Protocols 1(100) and 2(100):   0.1504E+04 0.0000E+00
 Protocols 1(100) and 3(100):   0.1305E+04 0.4128-279
 Protocols 2(100) and 3(100):   0.2592E+04 0.0000E+00

=== Doing statistics by repeating the experiment 100 times ===
-- Normalized to <A> = 1:
Protocol 1: <A> =  0.1000E+01, sigma =  0.5485E-02, /sqrt(S) =  0.5485E-03
Protocol 2: <A> =  0.9983E+00, sigma =  0.2008E-01, /sqrt(S) =  0.2008E-02
Protocol 3: <A> =  0.1036E+01, sigma =  0.2586E+00, /sqrt(S) =  0.2586E-01

--- chi-square test using 100 bins of length (4,2500) ---
              <khi2>     max(khi2)   P-value
 Protocol 1:  0.5044E+01  0.2388E+02  0.2287E-03
 Protocol 2:  0.4913E+04  0.2000E+05  0.0000E+00
 Protocol 3:  0.2587E+04  0.1201E+05  0.0000E+00

--- Fraction of events A(i,j) was selected ---
              (0,0)       (0,1)       (1,0)       (1,1)       (2,0)       (2,1)
 Protocol 1:  0.3146E-01  0.9388E-01  0.1240E+00  0.3755E+00  0.9440E-01  0.2808E+00
 Protocol 2:  0.3496E-01  0.1050E+00  0.1249E+00  0.3751E+00  0.9005E-01  0.2700E+00
 Protocol 3:  0.2737E-01  0.9676E-01  0.1100E+00  0.3914E+00  0.8259E-01  0.2918E+00
```